\documentclass{vldb}
\usepackage{graphicx}
\usepackage{subcaption}
\usepackage{algorithm}
\usepackage{algorithmic}
\usepackage{times,amsmath,epsfig,bm}
\usepackage{url}
\usepackage{mdwlist}
\usepackage{balance}
\usepackage{hyperref}
\usepackage{breakurl}
\usepackage{multirow}

\usepackage{color}

\begin{document}

\title{Community Specific Temporal Topic Discovery\\ from Social Media}

\numberofauthors{3}
\author{
\alignauthor
Zhiting Hu
\\
\affaddr{Department of Computer Science,Peking University} \\
\email{huzhiting@pku.edu.cn}
\alignauthor
Chong Wang \\
\affaddr{Machine Learning Department}\\
\affaddr{Carnegie Mellon University}\\
\email{chongw@cs.cmu.edu}
\alignauthor
Junjie Yao
\\
\affaddr{Department of Computer Science,University of California, Santa Barbara}\\
\email{jjyao@cs.ucsb.edu}
\and
\alignauthor
Eric Xing\\
\affaddr{Machine Learning Department}\\
\affaddr{Carnegie Mellon University}\\
\email{epxing@cs.cmu.edu}
\alignauthor
Hongzhi Yin\\
\affaddr{Department of Computer Science,Peking University}\\
\email{bestzhi@pku.edu.cn}
\alignauthor
Bin Cui\\
\affaddr{Department of Computer Science,Peking University} \\
\email{bin.cui@pku.edu.cn}
}
\maketitle

\begin{abstract}
Studying temporal dynamics of topics in social media is very useful to understand online user behaviors. Most of the existing work on this subject usually monitors the global trends, ignoring variation among communities. Since users from different communities tend to have varying tastes and interests, capturing community-level temporal change can improve the understanding and management of social content. Additionally, it can further facilitate the applications such as community discovery, temporal prediction and online marketing. However, this kind of extraction becomes challenging due to the
intricate interactions between community and topic, and intractable computational complexity.

In this paper, we take a unified solution towards the community-level topic dynamic extraction. A probabilistic model, CosTot~(Community Specific Topics-over-Time) is proposed to uncover the hidden topics and communities, as well as capture community-specific temporal dynamics.
Specifically, CosTot considers text, time, and network information simultaneously, and well discovers the interactions between community and topic over time.
We then discuss the approximate inference implementation to enable scalable computation of model parameters, especially for large social data. Based on this, the application layer support for multi-scale temporal analysis and community exploration is also investigated.

We conduct extensive experimental studies on a large real microblog dataset, and demonstrate the superiority of proposed model on tasks of time stamp prediction, link prediction and topic perplexity.







\end{abstract}

\section{Introduction}

With the prevalence  of online social networks, such as Twitter
and Facebook, social media has become a ubiquitous part of
people's daily lives. It provides a platform for users to post
short and quick-updated texts, exhibiting rich temporal dynamics.
Understanding these dynamics provides important insights into
people's changing online behaviors.  Extensive research is
devoted to uncover the temporal dynamics of online
content~\cite{lin2013voices, [Yangwsdm11], wang2006topics}.

\begin{figure}[!htp]
\centering
\begin{tabular}{c}
\includegraphics[width=0.8\columnwidth]{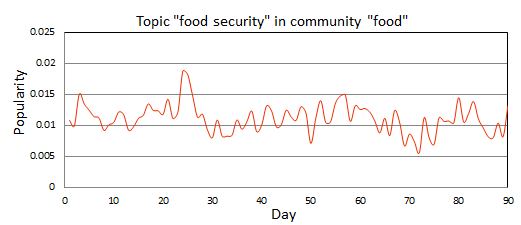}\\
\includegraphics[width=0.8\columnwidth]{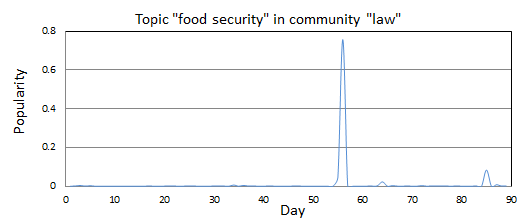}\\
\end{tabular}
\vspace{-15pt}
\caption{Temporal distributions of topic ``food security'' in
community ``food'' (top) and community ``law'' (bottom), over a
period of 90 days. We see different communities exhibit very
different patterns on the same topic.}
\label{fig:topic-foodsecurity}
\end{figure}

However, most of these existing work only explores global
temporal variation, or the overall trends of topics. 
This ignores an important aspect of social media---the {\it
communities}. A community is a collection of users with more
or/and better interactions amongst its members than the rest of
the global network~\cite{leskovec2010empirical}.
Communities play a crucial role in social media, and provide the
basis for user participation and engagement.
Members in the same community typically bear similar content
preferences and often communicate on shared
topics~\cite{sachan2012using,ruan2013efficient}.
Given that the content of social media is so dynamic, it is
expected that different communities tend to have different
temporal dynamics of topics.
One example from our experiments on the microblog data is
illustrated in Figure~\ref{fig:topic-foodsecurity}, where we show
the temporal distributions of topic ``food security'' in a
community interested in ``food'', and another community mainly
focusing on ``law'', respectively. We can clearly observe
different patterns. The huge burst in community ``law'' coincides
with the scandal of New Zealand substandard milk powder erupted
on Jan 25th, 2013.
By distinguishing the patterns of temporal variations across
different communities, we gain deeper insights on how topics
change over time, and how different pieces of content attract
attentions from different communities. This can be potentially
used for various applications.
For instance, in online marketing, advertisers are allowed to
achieve exact targeting for their advertisements with different
subjects, or design community-specific content to effectively
catch the eyes of community members.

Extracting community-specific temporal dynamics of topics can be
very challenging. First, user community membership is often
unknown, and the topics are also hidden. With the boom of both
links and text in social media, it is necessary to extract
community structures and latent topics simultaneously to attain a
complete view of the social media. However, even though there are
a bulk of studies on community detection and topic modeling
respectively, only a limited number of work~\cite{liu2009topic,
yan2012understanding} aims at jointly modeling these two
important aspects. Furthermore, the interactions between community and topic are not straightforward to model. For example, communities in
\cite{liu2009topic} do not have direct relations with topics, and
in \cite{yan2012understanding}
one community corresponds to only one topic, violating the fact
that a community in social media typically has varying degrees of
interests in different topics. Second, human activities are
highly volatile, especially on the online social networks where
diverse content grows and fades rapidly over time. Although the
aggregate temporal dynamics of individual behaviors may exhibit
certain patterns, different communities with various interests
tend to have their own temporal patterns. Third, the
unprecedented data scale, including enormous volume of text
and large sparse network, poses new computational challenges.


In this paper, we propose a probabilistic framework to address the above challenges. A probabilistic longitudinal model, CosTot (Community Specific Topics-over-Time) is developed
to uncover the hidden topics and
communities, as well as capture the community-specific temporal
topic variations, from large-scale social media data. Our model defines a generative process for text, time and network to accurately characterize social media. Specifically, our model assumes that 1) each user can belong to multiple communities with different degrees of affiliation strength; 
2) each community can have varying levels of interests in multiple topics; 3) each topic exhibits different temporal variations within different communities,
and these variations are determined by the content of user posts along the
time line.
Different from  existing temporal models such as Topics over Time
(TOT)~\cite{wang2006topics}, CosTot can provide a finer-grained
exploration of community-specific temporal dynamics of topics.
CosTot is also more accurate than~\cite{yan2012understanding} in terms of modeling the correlation between community and topic by allowing communities to
have mixture of topics rather than one-to-one correspondence.
To enable our proposed CosTot scalable to large-scale social media datasets,
we design a Gibbs Sampler: by implicitly modeling negative links in Bayesian prior, it takes, in each iteration, linear time in terms of the size of total words and number of positive links in the dataset. It can usually converge after a small number of iterations in our experiments.


We then deploy our approach to facilitate a set of applications on real microblog data. The patterns and knowledge learnt by CosTot enables the application layer to explore fine-grained topic temporal dynamics, analyze community temporal characteristics, as well as detect bursty events.

We further conduct extensive experiments on both synthetic and large-scale real datasets to evaluate the performance of our proposed method against state-of-the-art approaches. The real dataset consists of $11$M posts generated by $53$K users between 12/01/2012 to 02/28/2013. The results show the superiority of our method in terms of time prediction, link prediction and text perplexity, which indicates the advantage of CosTot in modeling temporal dynamics, network structure and text, respectively.

To summarize, we make the following contributions in our work.
\begin{enumerate}
\item We identify the problem of community-specific topic dynamic extraction from social media data with text, time and network. The rich features and interaction can be used in the framework of new setting. To the best of our knowledge, such a problem has not been investigated before.
\item We propose a unified probabilistic model, CosTot, which
  uncovers the topics and communities as well as captures the
  community-specific temporal dynamics of topics. We well study the features required for this model, and design an efficient inference algorithm to guarantee the scalability of our method.
\item We deploy our approach to facilitate a set of applications on real social media data. We also present a comprehensive study of community-specific topic temporal variations, and show its usefulness to social media analysis. 

\item We conduct extensive experiments to evaluate the performance of our approach on large-scale real dataset. The results show the superiority of our model in terms of modeling text, network structure and temporal dynamics.
\end{enumerate}





The rest of the paper is organized as follows: Section~\ref{sec:realted} reviews related literature;Section~\ref{sec:framework}
formulates the problem, and introduces the proposed model
and the inference algorithm; Section~\ref{sec:application} reveals the analysis and applications enabled by the proposed new model;
Section~\ref{sec:exp} presents our experimental results;
and finally we conclude this paper and outline future work.

\section{Related Work}\label{sec:realted}

In this section, we describe the related work in two areas:
temporal dynamics of topics and community detection.



%

\textbf{Temporal Dynamics of Topics}:
Topic models, such as latent Dirichlet
allocation (LDA)~\cite{blei2003latent} are usually utilized to
find latent topics from text collections. In topic models,
documents are modeled as a distribution over a shared set of
topics, while topics themselves are distributions over words.
Modeling temporal dynamics using topic models has attracted
huge interest. A number of temporal topic models were proposed.
Topics Over Time (TOT)~\cite{wang2006topics} models the text and time
stamp of a document jointly, assuming that latent topics generate
time stamp according to a Beta distribution. One main shortcoming
of Beta distribution is that it is unimodal, and thus limits the
available patterns of topic temporal variation. TOT can also be
seen as a special form of a more flexible model, supervised
latent Dirichlet allocation (sLDA)~\cite{mcauliffe2007supervised}.
Another set of approaches \cite{blei2006dynamic,
zhang2010evolutionary,
ren2008dynamic}
makes Markovian assumption on topic variation.
They divide time into epoches, and assume
that topics evolve based on their states in the
previous epoch,

In addition to topic models, there are also a bulk of
other approaches modeling temporal variation of information
diffusion. \cite{[Yangwsdm11]} develops the K-Spectral
Centroid (K-SC) clustering algorithm which finds six classes of
patterns of temporal variation. \cite{matsubara2012rise} proposes
SPICKM, a flexible analytical model that generalizes and
explains earlier theoretical models for the rise and fall patterns
of influence propagation.



A closely related line of work to temporal dynamics of topics is
bursty event detection~\cite{kleinberg2003bursty,
yao2010temporal, diao2012finding, yin2013AUnified}.
More specifically,
\cite{diao2012finding} applies a state machine-based method to
detect bursty topics discovered by TimeUserLDA~\cite{diao2012finding}.
\cite{yin2013AUnified}
simultaneously detects stable topics (e.g. topics on user
interest) and bursty topics (e.g.  topics on emergencies) in a
unified PLSA-based model.

Our work here is distinct from all the above methods as we not only
find the global trends of topics, but more importantly distinguish
patterns of temporal variation across different communities, and thus
provide a more thorough and fine-grained view for
temporal characteristics of social media. \cite{lin2013voices}
follows a similar line by tracking opinion shift of members from
two different groups, whereas a group is defined by some pre-defined
features and is far away from the network community as our work.


\textbf{Community Detection}: This has been a hot topic, especially in recent social community analysis~\cite{leskovec2010empirical}.
Numerous techniques have been developed to detect {\it disjoint} communities,
i.e., each user in the network is assigned to a single community.
However, in real life users are usually characterized by multiple
community memberships, leading to {\it overlapping}
communities. A review and comparative study of overlapping community
detection is presented by \cite{xie2011overlapping}.
Among all these work, blockmodeling~\cite{doreian2005generalized}
is based on statistical inference, and \cite{airoldi2008mixed}
introduces a mixed membership stochastic blockmodel in which
each user has a probability distribution over communities drawn from
a Dirichlet distribution. Different from our approach, these work uses network structures
alone to extract communities.

\begin{figure*}[h!tb]
\centering
\includegraphics[width=0.8\textwidth]{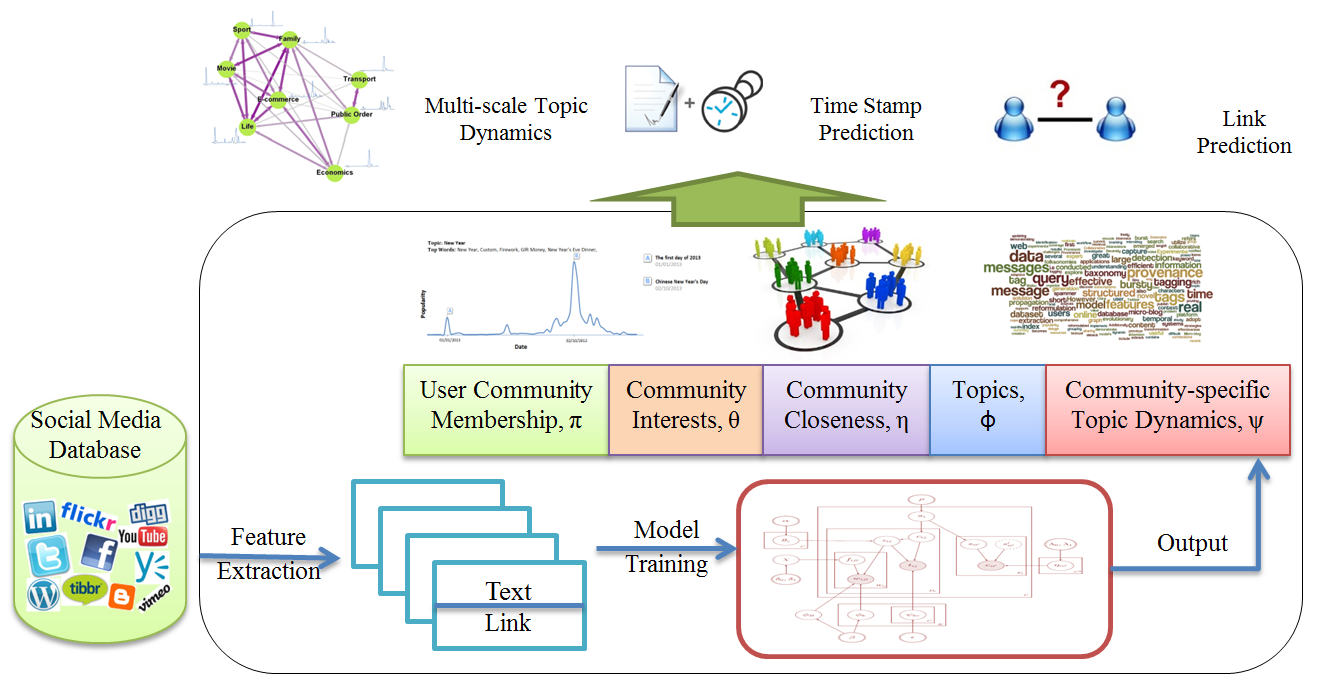}
\vspace{-13pt}
\caption{Framework of CosTot}
\label{fig:framework}
\end{figure*}

A growing number of recent papers incorporate both the network
structure and content to improve community detection performance.
For example, Topic-Link LDA~\cite{liu2009topic} jointly models
underlying topics of documents and author communities, and achieves
good performance in link prediction task. However, it does not uncover the relations between communities and topics.
There is also a line of work aiming at modeling documents and links between them (e.g. citations),
such as Pairwise-Link-LDA and Link-PLSA-LDA~\cite{nallapati2008joint},
RTM~\cite{chang2009relational} and PMTLM~\cite{zhu2013scalable}.

In these models, words and links are both generated by the same latent factor, which can be thought of as communities when generating links, and as topics when generating words. Hence there is a one-to-one correspondence between communities and topics. In contrast, our approach allows each community to have a mixture of topics, which is more reasonable due to the fact that communities in social media tend to have multiple interests.




\section{A Probabilistic Model for Community Specific Topic Dynamics} \label{sec:framework}

In this section, we introduce our framework for community-specific temporal dynamic discovery in social media, which can effectively support applications of temporal understanding and community analysis. We first formally define the problem we are interested in, then we propose a probabilistic model to uncover hidden topics and communities, as well as capture community-level topic temporal variations. Based on the model, we further design an efficient approximate inference algorithm.

%
\subsection{Problem Formulation}
We consider a social network $\mathcal{G}$ consisting of $U$
users. Each user $i \in \{1,\dots,U\}$ is associated with two
types of features.
\begin{enumerate}
  \item {\bf Text Data with Time Stamps:} a set of posts $\mathcal{D}_{i}=\{d_{i1}, d_{i2}, \\ \dots, d_{i|\mathcal{D}_{i}|}\}$ generated by user $i$. Each post $d_{ij} \in \mathcal{D}_{i}$ contains a bag of words from a given vocabulary, along with a time stamp $t_{ij}$, meaning that user $i$ generated post $d_{ij}$ at time $t_{ij}$.

  \item {\bf Network Data:} network links $\mathcal{E}_{i}=\{e_{ii'}|1\leq i'\leq U\}$ between $i$ and other users. Each link $e_{ii'} \in \mathcal{E}_{i}$ represents the social relationship between user $i$ and user $i'$. While we can adapt our model to describe either directed links or undirected links, we focus on directed links in this paper, since this is more common in microblog platforms such as twitter and weibo where $e_{ii'}\in\mathcal{E}_{i}$ means user $i$ follows user $i'$. We assume the links between users are constant within the time period we focus on. (We would like to relax this assumption in the future.)
\end{enumerate}

A {\it community} $c\in\{1,\dots,C\}$ has two components: a topic
probability vector $\theta_{c}$ where each component
$\theta_{ck}$ represents the probability that a post from the
community is related with corresponding topic $k$, and a link
formation probability vector $\eta_{c}$ where each component
$\eta_{cc'}$ is the probability that a user in community $c$
follows a user in community $c'$. Here $C$ is the total number of
communities.

%
Each user $i$ can belong to different communities (mixed
membership). That is to say, each user $i$ is associated with a
community probability vector $\pi_{i}$.

We define a {\it topic} $k\in\{1,\dots,K\}$ as a multinomial
distribution over the vocabulary, denoted as $\phi_{k}$. Here $K$
is the number of topics. For each
topic $k$, the {\it community-specific temporal dynamics} is
represented by a set of $C$ multinomial vectors
$\varphi_{k}=\{\varphi_{k1},\varphi_{k2},\dots,\varphi_{kC}\}$,
each of which represents the time variation of topic $k$ within
the corresponding community, i.e. a probability distribution over
discrete time slices. This kind of vector represents how the popularity or attention to topic $k$ in community $c$ changes over time.

Given the text data with time stamps and network data in social media, our goal is to
uncover hidden communities and topics, and infer the community-specific temporal dynamics of topics.
This extraction can improve
the understanding of information changes and community characteristics,
which further benefit several important
tasks in social media, e.g., link prediction, time stamp prediction and multi-scale dynamic analysis.

\subsection{Framework Overview}

In this paper, we propose a probabilistic framework to achieve these goals and support the upper layer applications.
The topics and communities are both hidden factors to be extracted, and the correlation between them are always omitted due to the modeling and inference complexity. Here we use a multiple stage approach to tackle this challenge. Specifically, our work can be listed as the following stages, and the framework is shown in Figure~\ref{fig:framework}.
\begin{enumerate}
\item \textbf{Feature Extraction:} We extract the text and network features of users from raw social media records. Here we discretize time line by dividing the time span of all users' posts into time slices.
     We then select the records from each consecutive time slots, and associate each post with corresponding time stamp.

\item \textbf{Model Training:} We introduce a probabilistic model to uncover the correlations between communities and topics over time. In this model, we combine the community extraction, topic identification, and community-specific topic dynamic discovery in a unified way. Though seeming complex, we tackle it with a well defined multiple component strategy and design an effective inference algorithm.
\item \textbf{Model Output:} We can get the temporal, topic and community information from the previous designed probabilistic model. We organize these intermediate output for later process.
\item \textbf{Dynamic Analysis:} The probabilistic model enables a set of novel applications based on the new data extraction. For example, we are allowed to gain a multi-scale view of topic temporal dynamics. In addition, by utilizing the fine-grained data representation, our approach further improves the extensively studied tasks of time and link prediction. We demonstrate how our approach supports these various applications in Section~\ref{sec:application}.
\end{enumerate}

For the model training part, we can enumerate the requirements and necessary steps towards this goal.
\begin{enumerate}
  \item For each user $i$, we need to infer the community probabilities $\pi_{i}$.
  \item For each community $c$, we are required to infer the topic probabilities $\theta_{c}$, and link probabilities $\eta_{c}$;
  \item For each topic $k$, infer the word probabilities $\phi_{k}$. This is necessary for topic representation.
  \item Infer the community-specific temporal dynamics $\varphi_{kc}$ of each topic $k$ within each community $c$. Based on the above three steps, we can unify them into the final stage. 
\end{enumerate}

\subsection{Community Specific Topics-Over-Time\\ Model}\label{subsec:costot}

Here we describe the proposed model, CosTot
(Community Specific Topics-over-Time), later we will show how to
perform inference with this model using a Gibbs sampling
algorithm which scales linearly with respect to the size of data.

CosTot is a latent space model jointly over text, time and network, and infers the patterns mentioned above.
Some of its building blocks are inspired by earlier
successful attempts, including the Mixed Membership Stochastic
Blockmodel (MMSB)~\cite{airoldi2008mixed} over networks, and
Topics over Time (TOT)~\cite{wang2006topics} over text and time.
Specifically, CosTot contains three closely linked components. The
user membership component models user membership to communities;
the network component explains the link structure; the text-time
component uncovers the semantic contexts, and captures the
temporal variations in different communities.  We will describe
each component in detail later in this section.

\begin{figure}[h!tp]
\centering
\includegraphics[width=0.8\columnwidth]{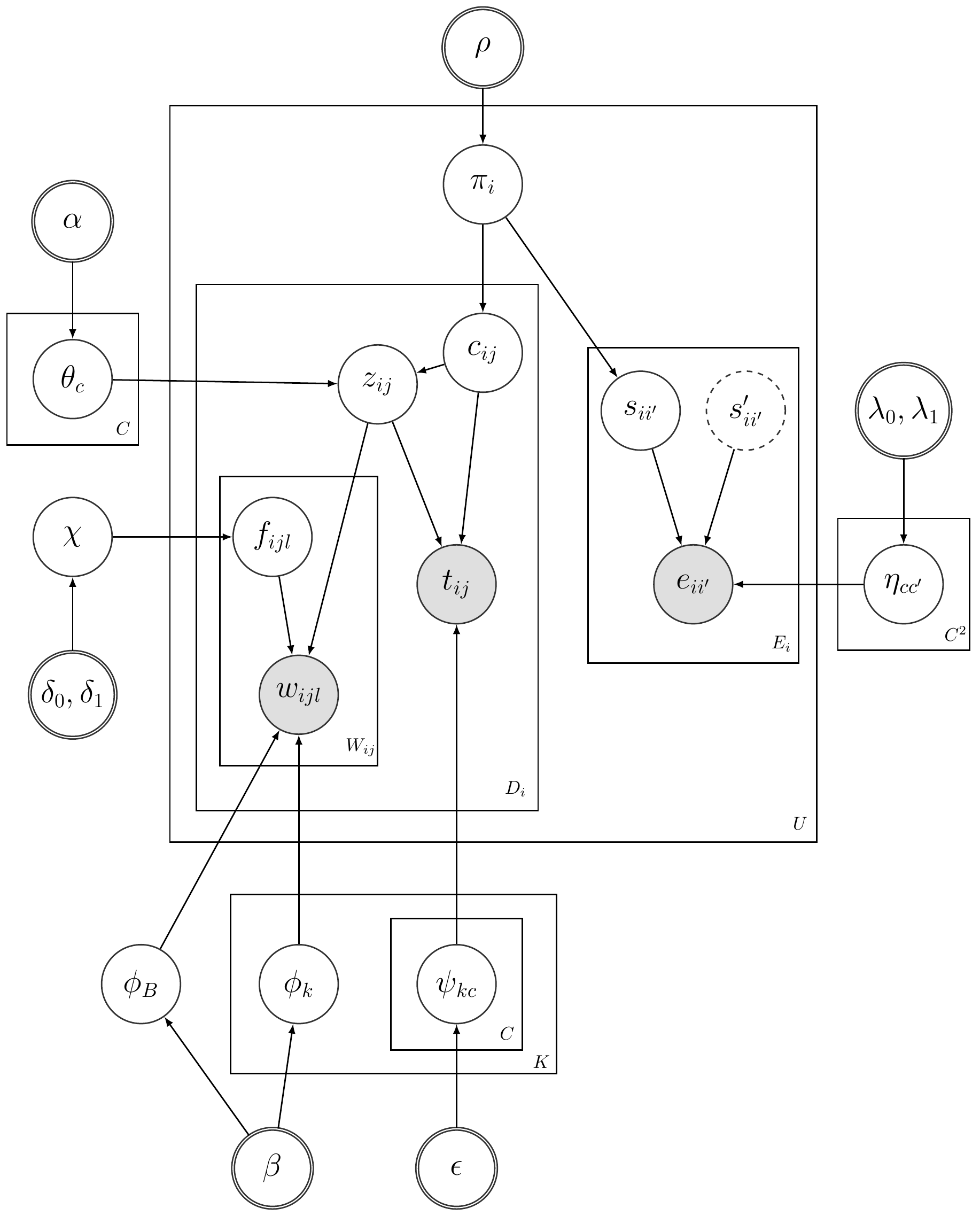}
\vspace{-10pt}
\caption{Graphical Model Representation of CosTot. A double circle indicates
a hyperparameter; a single hollow circle indicates a latent variable; and
a filled circle indicates an observed variable.
The latent variable $s'_{ii'}$ is represented as dashed circle since it is
drawn from $\pi_{i'}$ which is not shown in the graphical model.}
\label{fig:graphicalmodel}

\end{figure}

Our model is summarized as the generative process shown in Algorithm~\ref{algo:costot}.
Figure~\ref{fig:graphicalmodel} displays its graphical model representation.

\begin{algorithm}[h!tb]
\begin{enumerate*}
\item Sample foreground-background distribution,\\
     $\chi|\delta_{0},\delta_{1} \sim
  \text{Beta}(\delta_{0},\delta_{1})$.
\item Sample the background word topic, $\phi_{B}|\beta \sim \text{Dirichlet}(\beta)$.
\item For each topic $k = 1,2,\dots$, $K$,
    \begin{enumerate*}
    \item Sample the distribution over words, $\phi_{k}|\beta \sim \text{Dirichlet}(\beta)$.
    \item For each community $c = 1,2,\dots$, $C$,
         \begin{enumerate*}
         \item Sample the distribution over time stamps, $\psi_{kc}|\epsilon \sim \text{Dirichlet}(\epsilon)$.
         \end{enumerate*}
    \end{enumerate*}
\item For each community $c = 1,2,\dots$, $C$,
    \begin{enumerate*}
    \item Sample the distribution over topics, $\theta_{c}|\alpha \sim \text{Dirichlet}(\alpha)$.
    \item For each community $c' = 1,2,\dots$, $C$,
        \begin{enumerate*}
         \item Sample community-community link probability, \\
                $\eta_{cc'} | \lambda_{0}, \lambda_{1} \sim \text{Beta}(\lambda_{0},\lambda_{1})$.
        \end{enumerate*}
    \end{enumerate*}
\item For each user $i = 1,2,\dots, U$
    \begin{enumerate*}
    \item Sample the distribution over communities, \\
        $\pi_{i}|\rho \sim \text{Dirichlet}(\rho)$.
    \item For each post $j = 1,2,\dots$,
        \begin{enumerate*}
        \item Sample community indicator, $c_{ij}|\pi_{i} \sim \text{Multi}(\pi_{i})$.
        \item Sample topic indicator, $z_{ij}|\theta_{c_{ij}} \sim \text{Multi}(\theta_{c_{ij}})$.
        \item For each word $l = 1, 2, \dots$,
            \begin{enumerate*}
            \item Sample foreground indicator, $f_{ikl} \sim \text{Bernoulli}(\chi)$.
            \item Sample word, $w_{ijl}|\phi_{z_{ij}}
                    \sim \text{Multi}(\phi_{z_{ij}})$ if $f_{ikl}=1$,
                    or $w_{ijl}|\phi_{B}
                    \sim \text{Multi}(\phi_{B})$ if $f_{ikl}=0$.
            \end{enumerate*}
        \item Sample time stamp, $t_{ij}|\psi_{z_{ij}c_{ij}}
            \sim \text{Multi}(\psi_{z_{ij}c_{ij}})$.
        \end{enumerate*}
        \item For each user $i'=1, 2, \dots, U$ 
            \begin{enumerate*}
            \item Sample community indicator, $s_{ii'}|\pi_{i} \sim \text{Multi}(\pi_{i})$.
            \item Sample community indicator, $s'_{ii'}|\pi_{i'} \sim \text{Multi}(\pi_{i'})$.
            \item Sample link, $e_{ii'} | \eta_{s_{ii'}s'_{ii'}}
                    \sim
                    \text{Bernoulli}(\eta_{s_{ii'}s'_{ii'}})$.
            \end{enumerate*}
    \end{enumerate*}
\end{enumerate*}
\caption{Generative Process for CosTot}\label{algo:costot}
\end{algorithm}


{\bf User membership component.}\quad
Our proposed model uses a mixed membership approach~\cite{airoldi2008mixed}, to capture the fact that people bear multiple roles in social media and their behaviors
are influenced by different community context~\cite{xie2011overlapping}.
Note that in our model user membership integrates two aspects of user behaviors, i.e., posting (which generates text) and friending/following (which generates links). It is coherent with the observation that members in a community not only have denser links among each other than those from different communities, but also tend to be interested in similar topics.

For each user $i$, we would like to infer the probability that $i$ belongs to each of the $C$ communities. Hence $i$ is associated with a community
probability vector $\pi_{i}$.  Each post $d_{ij} \in
\mathcal{D}_{i}$ is assigned to a single community $c_{ij}$,
denoting the community membership of user $i$ when she writes the
post. In addition, each link $e_{ii'} \in \mathcal{E}_{i}$ is
associated with two communities $s_{ii'}$ and $s'_{ii'}$, one for
each of the two users $i$ and $i'$ respectively, denoting their
community memberships when user $i$ builds relationship with user
$i'$.

{\bf Text-time component.}\quad
Content of community $c$ is generated by a mixture $\theta_{c}$ of topics.
The distribution $\theta_{c}$ models community's varying levels of interests in multiple topics, and hence plays the critical role of connecting the two aspects of community and topic in social media.

Each post $d_{ij}\in\mathcal{D}_{i}$ contains a bag of words
$\{w_{ij1},\dots,w_{ij|d_{ij}|}\}$ where $|d_{ij}|$ denotes the
length of the post. In traditional topic models such as latent
Dirichlet allocation (LDA)~\cite{blei2003latent}, a document is
associated with a mixture of topics and each word has a
topic label. This is reasonable for long documents such as
academic papers. However, on social media like twitter or weibo,
a post is usually very short, and thus is most likely to be about
a single
topic~\cite{diao2012finding,zhao2011comparing,yan2012understanding}.
We therefore assume a single latent topic variable $z_{ij}$ with
$d_{ij}$ to indicate its topic.
In addition, posts are typically noisy, containing words
irrelevant to the main topics~\cite{zhao2011comparing,yan2012understanding}.
Hence, we assume a background word distribution $\phi_{B}$ to
capture such common words, and associate each word $w_{ijl}$ with
a background boolean indicator $f_{ijl}$ to indicate
 the word $w_{ijl}$ is
drawn from the background topic or not.

To model the discretized time stamps of posts, we
use a multinomial distribution $\varphi_{kc}$ over time stamps to model the time
variation specific to each topic $k$ and each community $c$.
Thus, a post $d_{ij}$ is generated at the time $t_{ij}$ drawn
from $\varphi_{z_{ij}c_{ij}}$. Compared to Topics over Time
(TOT)~\cite{wang2006topics} which uses a Beta distribution to
model time variations and only allows a unimodal distribution
over time for each topic, our use of multinomial distribution can
capture multimodal variations. It is more flexible in capturing
real-life topics which usually rise and fall for many times.

{\bf Network component.}\quad
Different from text data collected from individual
users, the network data is relational and thus violates the
classical independence or exchangeability
assumptions~\cite{airoldi2008mixed}. To address the problem, the
network component uses pairwise community Bernoulli distributions
to model the presence and absence of links between pairs of
users. For link $e_{ii'}$, the indicator $s_{ii'}$ and $s'_{ii'}$
denotes the community membership of user $i$ and user $i'$
respectively, when user $i$ built relationship with user $i'$.
Then $e_{ii'}$ is drawn from $\eta_{s_{ii'}s'_{ii'}}$ which
represents the relationship strength between community $s_{ii'}$
and $s'_{ii'}$.

The network of a social media is typically sparse, thus we only
model positive links: the variables $s_{ii'}, s'_{ii'}$ exist if
and only if $e_{ii'}\in \mathcal{E}_{i}$.  As in
\cite{yan2012understanding}, the negative links $e_{ii'} \not\in
\mathcal{E}_{i}$ are implicitly modeled in a Bayesian fashion: we
use a Beta($\lambda_{0}, \lambda_{1}$) prior on each
$\eta_{cc'}$, and set $\lambda_{0} = \text{ln}(n_{neg}/C^{2})$
and $\lambda_{1}=0.1$, where $n_{neg} = U(U-1) -
\sum_{i}|\mathcal{E}_{i}|$ is the number of negative links. In
this way, we reduce large amount of computation and achieve
linear complexity on network modeling, as explained in
Section~\ref{sect:time-complexity}.

\subsection{Approximate Inference Implementation}

We then proceed to propose a collapsed Gibbs sampler for
approximate inference of the CosTot model.  At each iteration of our Gibbs sampler,
we need to sample, for each post $d_{ij}$ by user $i$, both the
corresponding community indicator $c_{ij}$ and the topic
indicator $z_{ij}$, and for each link $e_{ii'}$ the corresponding
community indicators $s_{ii'}$ and $s'_{ii'}$. We are also required to
sample the per-word foreground indicator $f_{ijl}$. We discuss these separately.

{\bf Sampling community indicator $c_{ij}$ for post $d_{ij}$.}
We sample the community indicator $c_{ij}$ for post $d_{ij}$
according to , \begin{equation}\label{eq:doc-comm}
\begin{split}
 &P(c_{ij} = c | z_{ij}=k, t_{ij}=t,\mbox{\boldmath$c$}_{-ij}, \mbox{\boldmath$s$},
 \mbox{\boldmath$z$}_{-ij}, \mbox{\boldmath$t$}_{-ij},.)\\
 &\propto \frac{n^{(c)}_{i}+\rho}{n^{(\cdot)}_{i}+C\rho}\cdot
 \frac{n^{(k)}_{c}+\alpha}{n^{(\cdot)}_{c}+K\alpha}\cdot
 \frac{n^{(t)}_{ck}+\epsilon}{n^{(.)}_{ck}+T\epsilon},
\end{split}
\end{equation}
where $n^{(c)}_{i}$ denotes the number of posts and links of user $i$ generated
by community $c$; $n^{(\cdot)}_{i}$ is the total number of posts and links of
user $i$; $n^{(k)}_{c}$ is the number of posts assigned to community $c$
and generated by topic $k$; $n^{(\cdot)}_{c}$ is the total number of posts generated
by community $c$; $n^{(t)}_{ck}$ denotes the number of times that
timestamp $t$ is generated by community $c$ and topic $k$, and
$n^{(\cdot)}_{ck}$ denotes the total number of timestamps generated by
community $c$ and topic $k$. All the counters mentioned above are calculated with
the post $d_{ij}$ excluded.

{\bf Sampling community indicators $s_{ii'}$ and $s'_{ii'}$ for link
$e_{ii'}$.}
Recall that we only model $s_{ii'}$, $s'_{ii'}$ and
$e_{ii'}$ for positive links $e_{ii'}\in\mathcal{E}_{i}$.  The
resulting conditional posterior distribution is:
\begin{equation}\label{eq:link-comm}
\begin{split}
 &P(s_{ii'} = c, s'_{ii'} = c'| \mbox{\boldmath$s$}_{-ii'}, \mbox{\boldmath$c$},
 \mbox{\boldmath$e$}, .)\\
 &\propto \frac{n^{(c)}_{i}+\rho}{n^{(\cdot)}_{i}+C\rho}\cdot
 \frac{n^{(c')}_{i'}+\rho}{n^{(\cdot)}_{i'}+C\rho}\cdot
 \frac{n_{cc'}+\lambda_{1}}{n_{cc'}+\lambda_{0}+\lambda_{1}},
\end{split}
\end{equation}
where $n_{cc'}$ is the number of positive links, with $e_{ii'}$ excluded,
whose communities indicators are $\{c, c'\}$.

{\bf Sampling topic indicator $z_{ij}$ for post $d_{ij}$}.
This is done through the conditional posterior probability
\begin{equation}\label{eq:doc-topic}
\begin{split}
\!\!\!\! &P(z_{ij} = k | c_{ij} = c, t_{ij} = t, \mbox{\boldmath$c$}_{-ij},
 \mbox{\boldmath$z$}_{-ij}, \mbox{\boldmath$f$}, \mbox{\boldmath$w$}, \mbox{\boldmath$t$}, .) \\
 \!\!\!\! &\propto \frac{n^{(k)}_{c}+\alpha}{n^{(\cdot)}_{c}+K\alpha}\cdot
 \frac{n^{(t)}_{ck}+\epsilon}{n^{(.)}_{ck}+T\epsilon}\cdot
 \frac{\prod^{V}_{v=1}\prod^{n^{(v)}_{ij}-1}_{q=0}(n^{(v)}_{k}+q+\beta)}
      {\prod^{n^{(\cdot)}_{ij}-1}_{q=0}(n^{(\cdot)}_{k}+q+V\beta)},
\end{split}
\end{equation}
where $n^{(v)}_{ij}$ is the number of times word $v$ occurs in the post $d_{ij}$ and is labeled
as a foreground word; $n^{(\cdot)}_{ij}$ is the total number of foreground words in the
post $d_{ij}$; $n^{(v)}_{k}$ denotes the number of times word $v$ is assigned to topic $k$,
and $n^{(\cdot)}_{k}$ is the total number of words assigned to topic $k$. Note that
$n^{(v)}_{k}$ and $n^{(\cdot)}_{k}$ are calculated with the post $d_{ij}$ excluded.

{\bf Sampling foreground indicator $f_{ijl}$ for word $w_{ijl}$}.
The conditional posterior distributions for the foreground indicator
$f_{ijl}$ are,
\begin{equation}\label{eq:word-foreback-1}
\begin{split}
  &P(f_{ijl} = 1 | z_{ij}=k, w_{ijl} = v, \mbox{\boldmath$z$}_{-ij},
  \mbox{\boldmath$w$}_{-ijl}, .)\\
  &\propto \frac{n^{(1)}+\delta_{1}}{n^{(\cdot)}+\delta_{0}+\delta_{1}}\cdot
  \frac{n^{(v)}_{k}+\beta}{n^{(\cdot)}_{k}+V\beta},
\end{split}
\end{equation}
and
\begin{equation}\label{eq:word-foreback-0}
\begin{split}
  &P(f_{ijl} = 0 | w_{ijl} = v, \mbox{\boldmath$z$},
  \mbox{\boldmath$w$}_{-ijl}, .)\\
  &\propto \frac{n^{(0)}+\delta_{0}}{n^{(\cdot)}+\delta_{0}+\delta_{1}}\cdot
  \frac{n^{(v)}_{B}+\beta}{n^{(\cdot)}_{B}+V\beta},
\end{split}
\end{equation}
where
$n^{(0)}$ and $n^{(1)}$ are the number of
background words and foreground words respectively; $n^{(\cdot)}
= n^{(0)} + n^{(1)}$; $n^{(v)}_{B}$ denotes the number of times
word $v$ is generated by background topic, and $n^{(\cdot)}_{B}$
is the total number of words generated by background topic.
Again, all the above counters are calculated with word $w_{ijl}$
excluded.

%
%
%

\subsubsection{Linear Time Complexity} \label{sect:time-complexity}

After the inference description, here we analyze the time complexity of this inference
algorithm. It is shown that the chosen algorithm scales linearly in terms
of the size of data, i.e. the number of words and positive
links, achieving satisfying performance.

Be ware that, in each iteration, the communities of each user's posts are first
sampled. Since all the counters (e.g. $n^{(c)}_{i}$) involved in
Eq.(\ref{eq:doc-comm}) can be cached and updated in constant time
for each $c_{ij}$ being sampled, Eq.(\ref{eq:doc-comm}) can be
calculated in constant time. Thus, sampling all
$\mbox{\boldmath$c$}$ takes linear time w.r.t the number of
posts. Next, we sample community indicators $\mbox{\boldmath$s$}$ using
Eq.(\ref{eq:link-comm}). Since we have implicitly modeled negative links in Bayesian piror (i.e., the Beta prior for $\eta_{cc'}$), we only need to sample $s_{ii'}$ and $s'_{ii'}$ for positive links $e_{ii'}\in\mathcal{E}_{i}$. Hence the complexity is reduced from quadratic (w.r.t the number of users) to linear (w.r.t the number of links). It significantly saves computation cost due to the sparseness of networks.
Finally, sampling all $\mbox{\boldmath$z$}$ and $\mbox{\boldmath$f$}$ by
Eq.(\ref{eq:doc-topic}), (\ref{eq:word-foreback-1}) and
{(\ref{eq:word-foreback-0})} is linear in the number of words.
Overall the inference algorithm takes linear time in the amount
of data.



\section{Applications Based on CosTot}\label{sec:application}

In this section, we demonstrate the usefulness of our approach by various representative applications on real-world social media data. We show that CosTot can effectively uncover communities and topics, and simultaneously capture temporal dynamics of topics in different communities. Based upon the patterns inferred by our model, we are allowed to explore topic dynamics at multiple granularities, identify bursty events, and give an in-depth analysis of particular communities. Our method can also support time stamp prediction and link prediction. 

\textbf{Real Data Setting:} We first introduce the data set used in the following study.
Our data is crawled from Sina Weibo\footnote{http://weibo.com},
one of the most popular microblog platforms in China. After
removing stop words and low active users with fewer than 20
posts, we obtain our dataset consisting of about $53$K
users, $2.1$M links, $11$M posts and $91$M words with a
vocabulary of size $89$K. The posts are distributed evenly in the
time period from December 1st 2012 through February 28th 2013.
Each post is labeled by the date it was posted. Therefore, time
stamps of the dataset range from 1 to 90. 

\subsection{Multi-scale Topic Temporal Dynamics}


In addition to most of existing works that only captures global
trends of topics, CosTot can detect community-specific trends
of topics. This allows us to have a multi-scale view of temporal dynamics,
as well as gain deeper insight on how topics
attract attentions from different communities.

\begin{figure}[!htp]
\centering
\includegraphics[width=1.05\columnwidth]{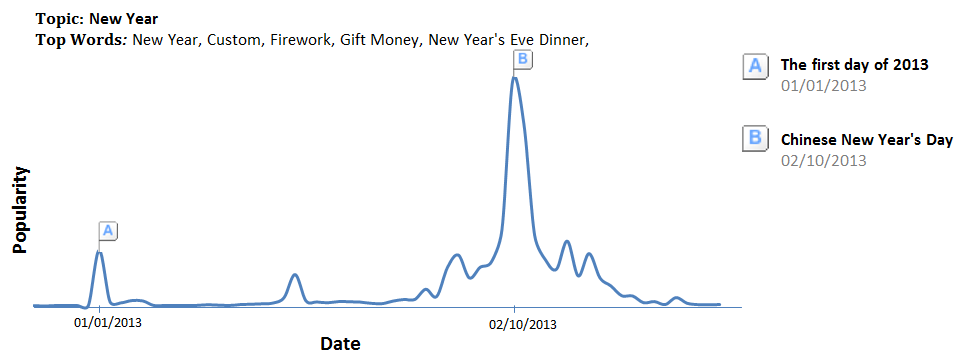}
\vspace{-18pt}
\caption{Global temporal dynamic. The ``Top Words'' on the top left
shows the top five most probable words in the topic,
based on which we give the topic a concise name ``New Year''.
Popular peaks are labeled with lettered flags. By manually
examining the data, we give a brief explanation for each peak to
the right of the timeline. \label{fig:whole-final-a}}
\end{figure}

Figure~\ref{fig:whole-final-a} shows the global trends of topic
``New Year'' in Sina Weibo, where the global popularity of
topic $k$ at time stamp $t$ is obtained by summing over all
communities (denoted as $c$) and users (denoted as $i$),
\begin{equation*}
\begin{split}
P(t | k) &= \sum_{c}P(t | c, k)P(c | k) \\
  &\propto \sum_{c}P(t | \psi_{kc})P(k | \theta_{c})\sum_{i}P(c|\pi_{i}),
\end{split}
\end{equation*}
where we assume the prior distribution $P(i)$ of each user $i$ is
constant.
We label the spikes of timeline with lettered flags. To better
understanding the semantic context of the spikes, we manually
check the posts generated around corresponding time stamps and
give a brief explanation for each spike. For example, the spike
``B'' on Feb 10th, 2013 is the Chinese New Year's Day,
while the lower spike ``A'' on Jan 1st, 2013 corresponds
to the first day of 2013. (Note our data was from Sina Weibo, a
Chinese website.)

\begin{figure}[!htp]
\centering
\includegraphics[width=1.0\columnwidth]{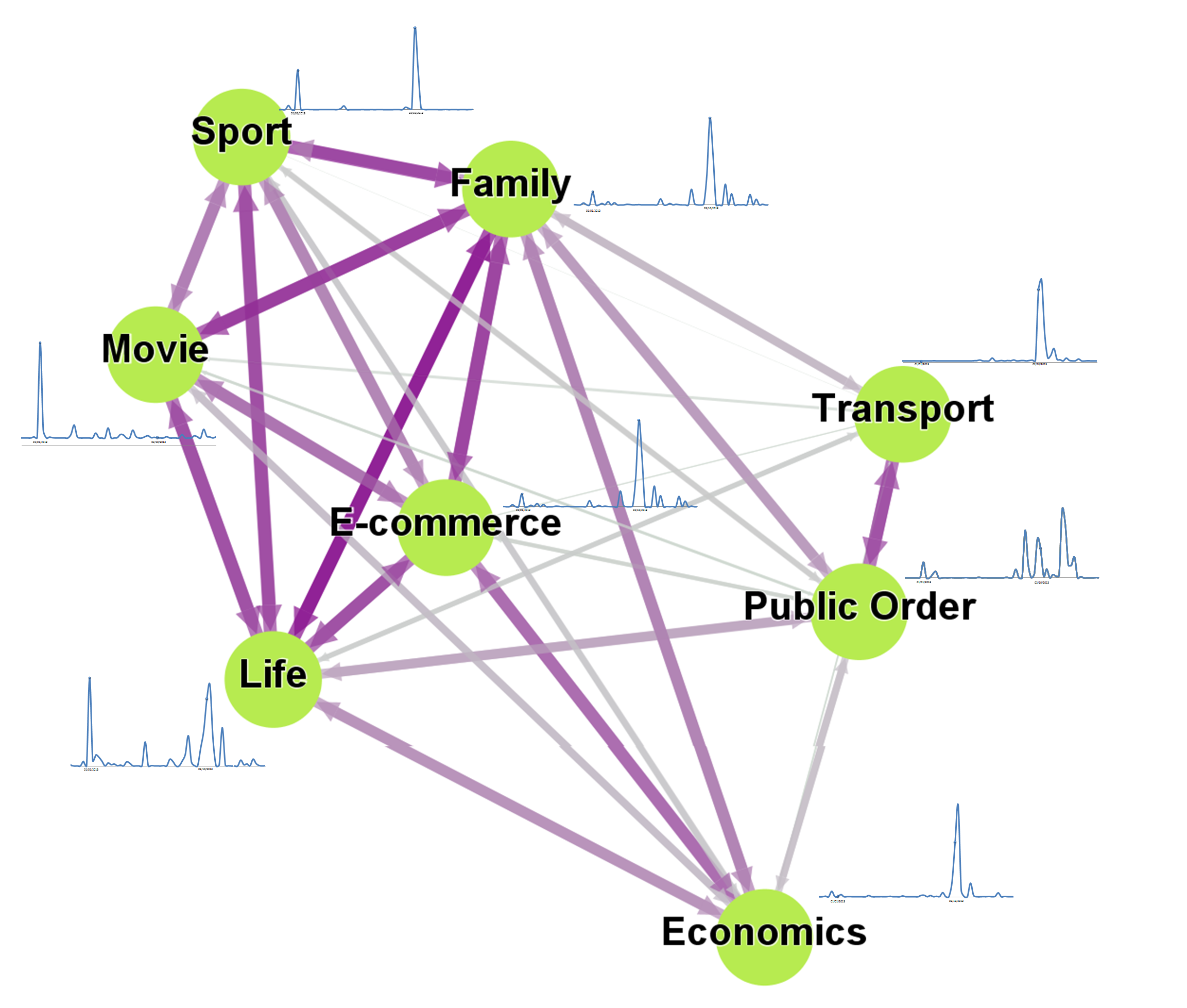}
\vspace{-13pt}
\caption{Community-specific temporal
dynamics. The global and community-specific temporal dynamics of
topic ``New Year''.  We focus on the time period starting on
12/25/2012 since the popularity is zero before that time. \label{fig:whole-final-b}}
\end{figure}

Figure~\ref{fig:whole-final-b} shows part of
the communities in Sina Weibo, where each node represents a
community. We choose a label for each community based on its
distribution over topics.  For example, community ``Movie'' puts
high probabilities on movie-related topics~\footnote{\footnotesize It is worth
mentioning that, in the results inferred by our model,
there is no community whose topic probability vector is dominated by topic
``New Year''. This is consistent with the fact that New Year is a temporary
event and unlikely to be some community's major interest.}.

The darkness of the
color of edges indicates the link probability between
corresponding communities---the darker the color is, the higher the
probability. For instance, there is a probability of $0.7$ that
the users in community ``Family'' follow the users in community ``Life'',
while the probability that the users in community ``Transport''
follow the users in community ``'Movie'' only achieves $0.2$. For
comparison, we put the community-specific dynamics of topic
``New Year'' near the corresponding nodes (communities). We see
that while most of the timelines peak around the Chinese New
Year's Day (i.e. spike ``B'' in Figure~\ref{fig:whole-final-a}),
only those communities on the left (e.g. community
``Movie'') have spikes around the first day of 2013 (i.e. spike
``A'' in Figure~\ref{fig:whole-final-b}), and communities on the
right (e.g. community ``Transport'') pay little attention to
the topic at that time. One possible explanation for the
phenomenon is that the communities on the left are more relevant to
entertainment, while those on the right seem to be more
concerned with professional stuff. Furthermore, strong links
between communities tend to make the temporal dynamics in
corresponding communities more similar, because they provide more
effective channels for information diffusion. The
community-specific temporal dynamics also suggest that the huge
spike ``B'' in the global dynamic is contributed by most communities
of the social media, and by contrast, the small spike ``A''
is formed due to the attentions from part of the communities.



\subsection{Characters of Extracted Communities}
Here we demonstrate that, by focusing on community-specific results, CosTot model enables us to step into a finer granularity and
get in-depth characteristics of particular
communities.

\begin{figure}[h!tb]
\centering
\includegraphics[width=1.0\columnwidth]{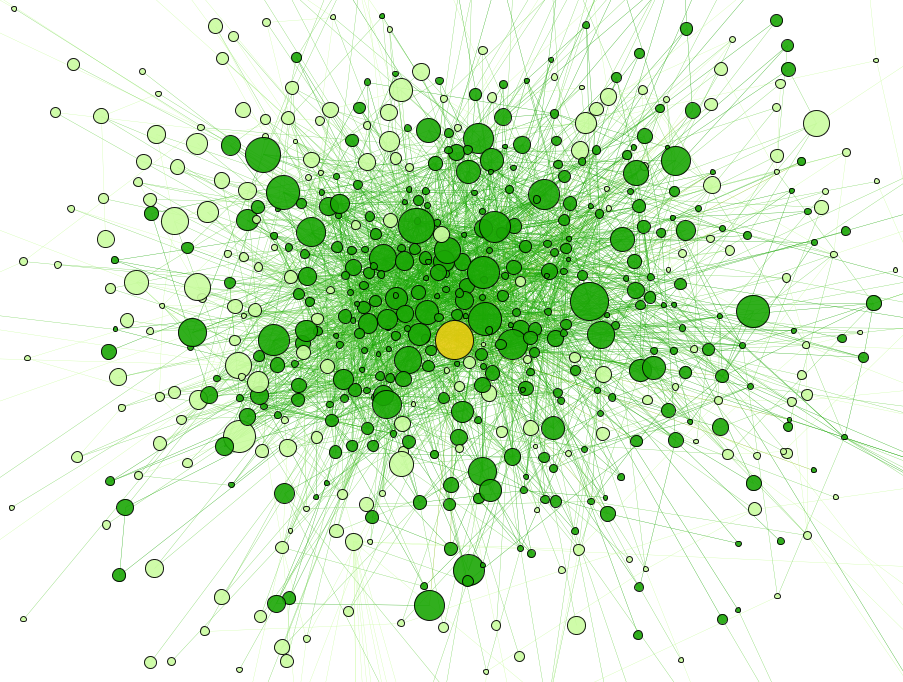}
\caption{Nodes and Temporal Change in Community ``E-commerce''. Size of each node is proportional to
the user's contribution (see text for more details). The node in yellow represents
the central actor of the community. Nodes that can reach the central actor within 2
steps along the directed edges are highlighted in dark green.}\label{fig:community-ecommerce}
\end{figure}

{\bf Community connectivities and user contributions.}
Figure~\ref{fig:community-ecommerce} provides a visualization for
community ``E-commerce''. Since our model captures mixed
membership of users, we define the \textit{contribution} of a
user $i$ to a community $c$ by considering both the membership
probability and number of posts generated by $i$, as
\begin{equation*}
\begin{split}
\text{contribution(i, c)} = \pi_{ic}\cdot\text{log}|\mathcal{D}_{i}|.
\end{split}
\end{equation*}
For community ``E-commerce'', we calculate contributions of all
users, and find that the contribution distribution approximately
follows a classic power law. Due to the space constraint, we omit this distribution figure. We select top users with    
contributions larger than $1.0$, which yields a subset of about $750$
users (Figure~\ref{fig:community-ecommerce}). The size of each
node is proportional to the user's contribution. We further recognize the
\textit{central actor}~\cite{wasserman1994social} of the
community as the user with highest contribution (the yellow
node), and as in~\cite{pathak2008social}, we highlight the nodes
that can reach the central actor within 2 steps along the
directed edges (i.e. ``following'' relationship).  From the
figure we see that most of the nodes are highlighted,
suggesting that the members are closely connected.

{\bf Topic dynamics within the same community.}

\begin{figure}[h!tb]
\centering
\includegraphics[width=1.0\columnwidth]{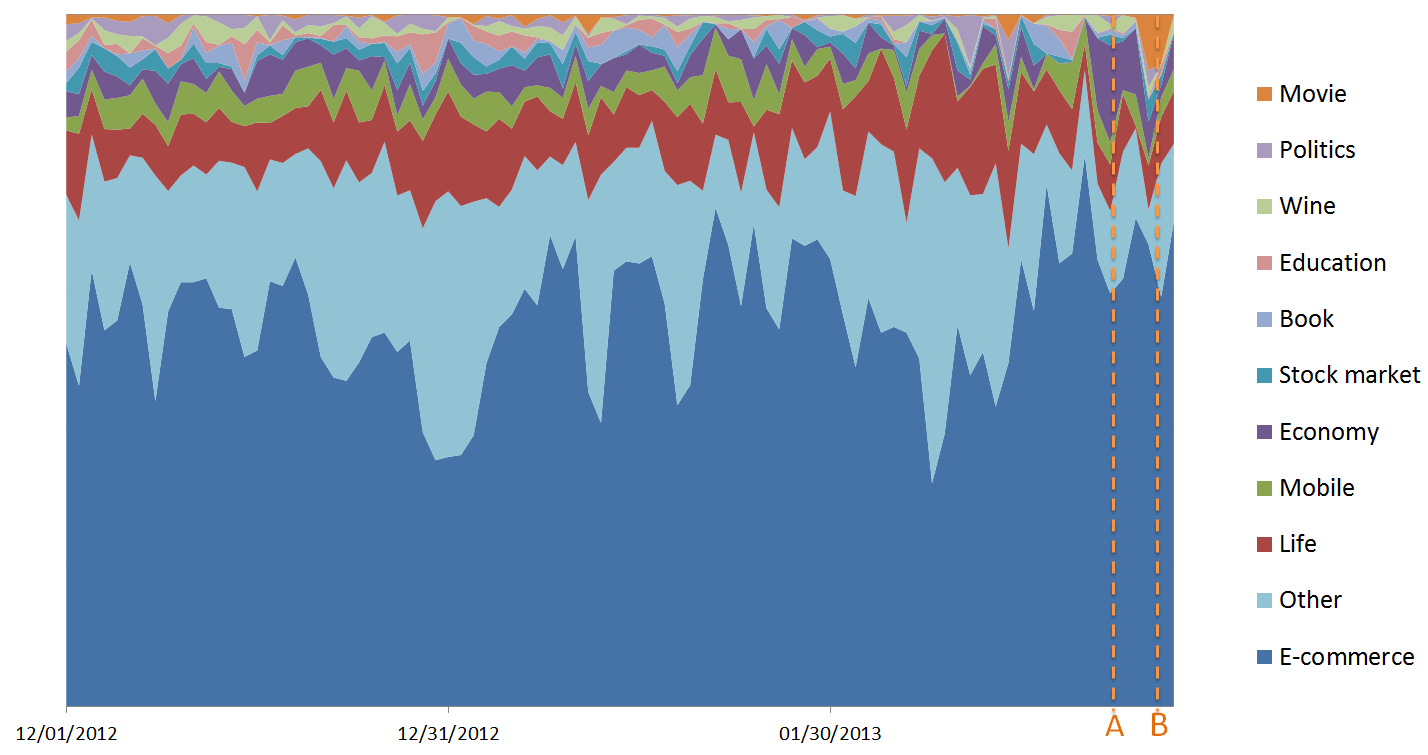}
\vspace{-13pt}
\caption{The distribution over topics as a function of time .
The legend shows the name for each topic and its corresponding color.
Top $10$ topics w.r.t community-topic distribution are plotted, while ``Other'' represents
the remaining $90$ topics.}\label{fig:stack-plot}
\end{figure}

To give a holistic view of the temporal dynamics of the
community's attention to different topics, we compute the
community's distribution over topics given time with the Bayes
rule:
\begin{equation*}
\begin{split}
P(k | t, c) = \frac{P(t | \psi_{kc})P(k | \theta_{c})}{\sum_{k}P(t | \psi_{kc})P(k | \theta_{c})}.
\end{split}
\end{equation*}
Figure~\ref{fig:stack-plot} shows the resulting patterns. The
height of a topic's region indicates the relative popularity of
the topic in the community ``E-commerce'' at given time. We can
observe that the attentions of the community members change over
time, while topic ``E-commerce'' dominates the focus. (That is
why we name the community as ``E-commerce''.) It is also notable
that other minor topics are also competing for the attention from
the community. For example, at time stamp $A$, topic ``Economy''
gains more concern among these minor topics, while afterwards (at
time stamp $B$) topic ``Movie'' successfully catches the eyes of
the members.


\begin{figure}[h!tb]
\centering
\subcaptionbox{Popularity of topic ``Life'' peaked when the new year is drawing near.}
{\includegraphics[width=1.0\columnwidth]{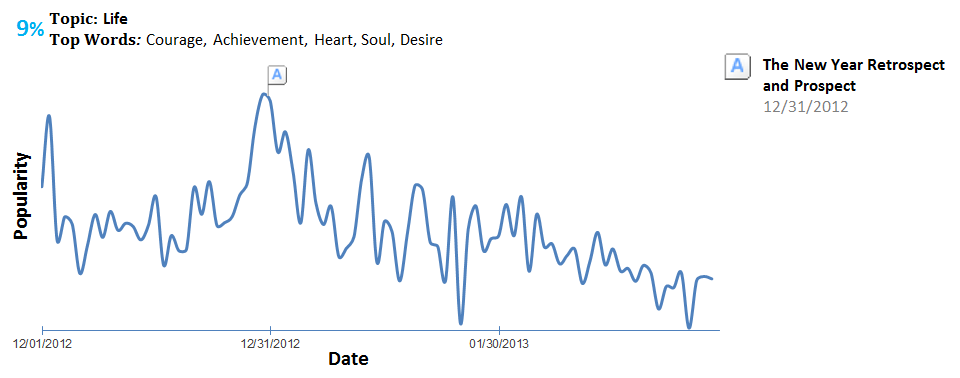}}

\subcaptionbox{Movie-related events significantly drew attentions of the community.}
{\includegraphics[width=1.0\columnwidth]{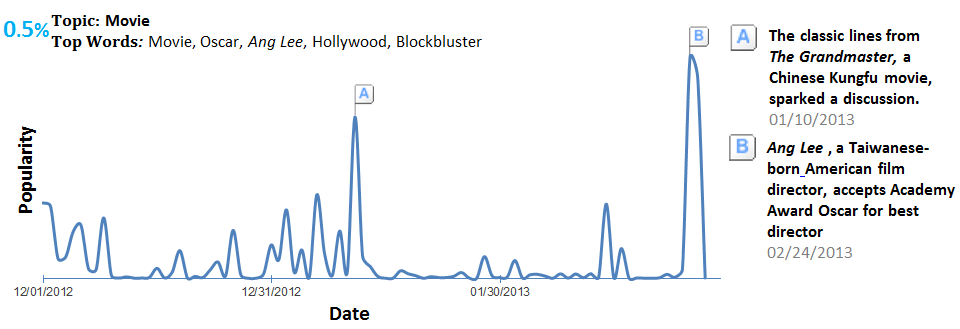}}
\vspace{-8pt}
\caption{Community ``E-commerce''-specific temporal dynamics of two topics with various levels of interest from the community. }
\label{fig:ecommerce}
\end{figure}

It is also interesting to step further by focusing on
individual topics and analyzing their temporal dynamics in the
particular community. We select two topics and display their
variation within community ``E-commerce'' in
Figure~\ref{fig:ecommerce}.  We find that it is {\it Ang Lee's}
win for best director in the {\it Academy Award} that boosts
topic ``Movie'' at time stamp $B$ in Figure~\ref{fig:stack-plot}.
It can also be observed that timelines of
topics with higher levels of interest tend to be smoother, while
topics with low levels of interest receive spiky attentions from
the members. The result indicates that temporal
dynamics of topics in ``unrelated'' communities can facilitate
bursty event detection. 
We further verify this claim in the next section.

\subsection{Event Identification}
\label{sect:event-detection}

Bursty event detection~\cite{kleinberg2003bursty,
diao2012finding} aims to capture the most popular events that
have drawn the public's attentions. When we want to find such
events related to a certain topic, one may intuitively focus on
the community whose major interests lie in the particular
topic, and analyze the behaviors of its members to detect bursts.
Our results, however, suggest that analyzing communities that are
not regularly concerned with the topic may provide a easier
way for bursty event detection.
\begin{figure}[h!tb]
\centering
\subcaptionbox{Topic ``Sports'' in community ``Sports''}
{
\includegraphics[width=1.0\columnwidth]{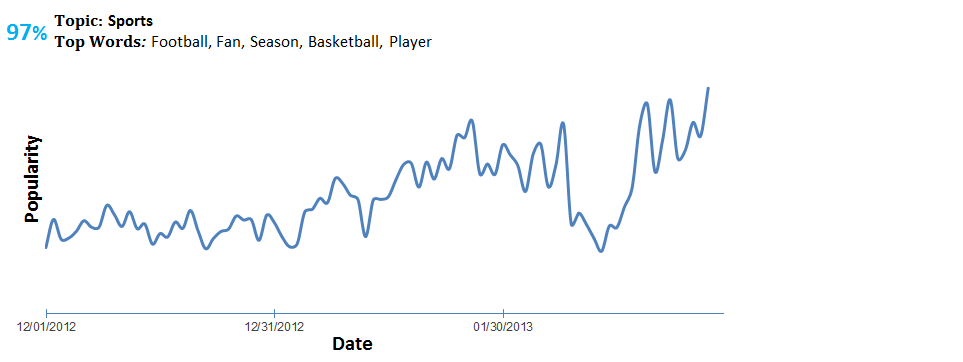}}\\
\subcaptionbox{Topic ``Sports'' in community ``Movie''}{
\includegraphics[width=1.0\columnwidth]{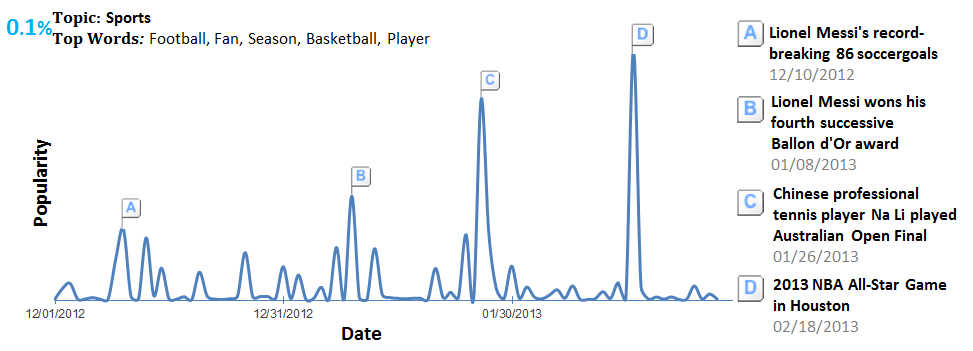}}\\
\vspace{-8pt}
\caption{Temporal dynamics of Topic ``Sports'' in Different Communities. Bursty events can be easily
identified in (b).}
\label{fig:topic-sports-final}
\end{figure}


An example is shown in Figure~\ref{fig:topic-sports-final}, where
(a) demonstrates the temporal dynamic of topic ``Sports'' in
community ``Sports'' (i.e. the community takes major interest in
topic ``Sports''), whereas (b) is the temporal dynamic of the
same topic in community ``Movie''. The percentages shown on the
top left corner represent the topic probability of corresponding
communities. From Figure~\ref{fig:topic-sports-final}(b) we can
more easily identify the bursts, such as the burst ``C'' which
coincides with the Australian Open Final the famous Chinese
professional tennis player Na Li attended.  On the contrary, the
timeline in Figure~\ref{fig:topic-sports-final}(a) is much
smoother and without clear spikes. By examining the data, we
found that community ``Sports'' did talk about the particular
events detected by Figure~\ref{fig:topic-sports-final}(b) when
they happened. However, these bursty behaviors are concealed
since members keep talking about sports throughout the time
period, and this results in no clear spikes. On the other hand,
members in community ``Movie'' are not concerned with
sports-related topic in the daily life, but their attentions
would still be drawn when significant events happened.  It is
worth mentioning that although the probability that community
``Movie'' generates sports-related posts is low, there are still
many members and large number of such posts involved due to the
large size of the data. Thus, the spikes represent the attentions
from this whole community, rather than from a small portion of its
members.

\subsection{Time Stamp/Link Prediction}\label{sec:app-pred}

\textbf{Time Stamp Prediction. }Another application of modeling temporal dynamics is to predict the time stamp of
a previously unseen document based on its content. It can be used to recover time stamps of documents with missing or incorrect meta-data~\cite{walker2012topics}.

Given the words in a post and its author, we
predict its time stamp by choosing the one that gives maximum
likelihood. Specifically, for a post $d$ by user $i$, its predicted time stamp is:
\begin{equation*}
\begin{split}
\hat{t}_{d} = \arg\max_{t}&\sum_{c}P(c|\pi_{i})\sum_{k}P(k|\theta_{c})P(t|\psi_{kc})\\
&\prod_{l}(\chi P(w_{dl}|\phi_{k})+(1-\chi)P(w_{dl}|\phi_{B})).
\end{split}
\end{equation*}

\textbf{Link Prediction. }Our framework also supports to predict the probability of a link between two users. Link prediction in social media not only helps in analyzing networks with missing data, but also can be used to recommend friends or followees by identifying very likely but not yet existent links~\cite{lu2011link}.

Based on the results of CosTot, the link prediction algorithm works as follows: for a pair of users ($i$, $i'$), we compute the probability of a link from user $i$ to $i'$ as
\begin{equation*}
\begin{split}
P_{i\to i'} = \sum_{s}\sum_{s'}P(s|\pi_{i})P(s'|\pi_{i'})\eta_{ss'},
\end{split}
\end{equation*}
and predict that the link exists if $P_{i\to i'}$ exceeds some threshold.

We will present the empirical improvements of time stamp and link prediction tasks in the later experimental study.


%
%
%
%

\section{Experiments}\label{sec:exp}

In this section, we conduct extensive experiments on both synthetic and real-world data to evaluate our proposed approach. For synthetic dataset, our experimental results show that CosTot is able to precisely uncover the hidden patterns. For real-world dataset, we demonstrate the superiority of our method by comparing it with state-of-the-art methods in various aspects.

\subsection{Experiments on Synthetic Data}



\begin{figure}[h!tb]
\centering
\begin{tabular}{cc}
\includegraphics[width=0.48\columnwidth]{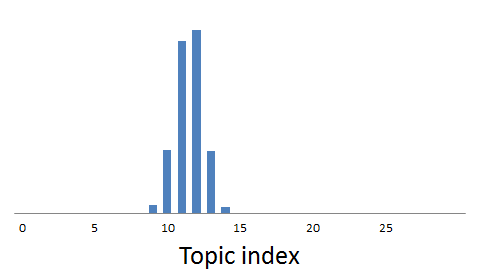}
&
\includegraphics[width=0.48\columnwidth]{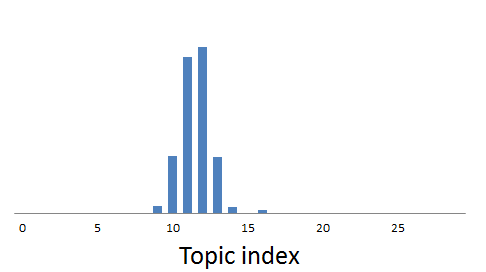}
\\
{\small (a.1)} & {\small (a.2)}
\\
\includegraphics[width=0.46\columnwidth]{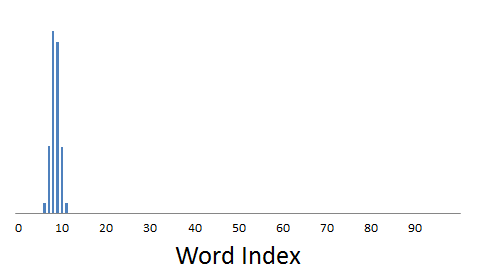}
&
\includegraphics[width=0.46\columnwidth]{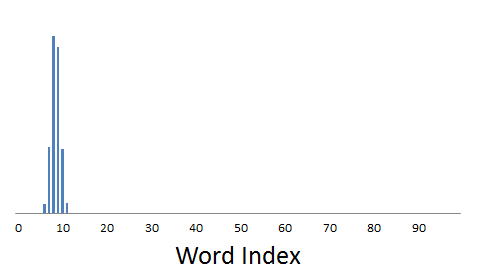}
\\
{\small (b.1)} & {\small (b.2)}
\\
\includegraphics[width=0.46\columnwidth]{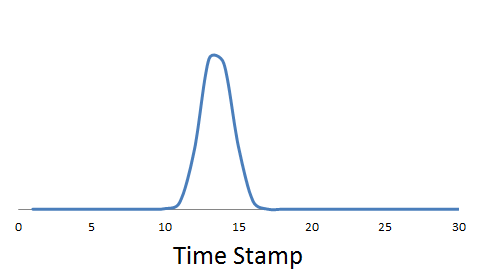}
&
\includegraphics[width=0.46\columnwidth]{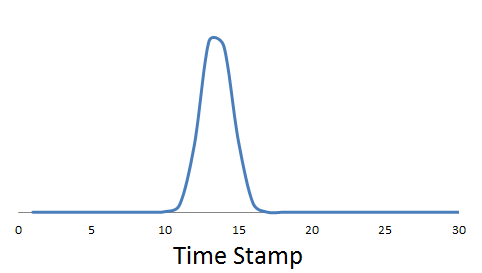}
\\
{\small (c.1)} & {\small (c.2)}
\\
\includegraphics[width=0.46\columnwidth]{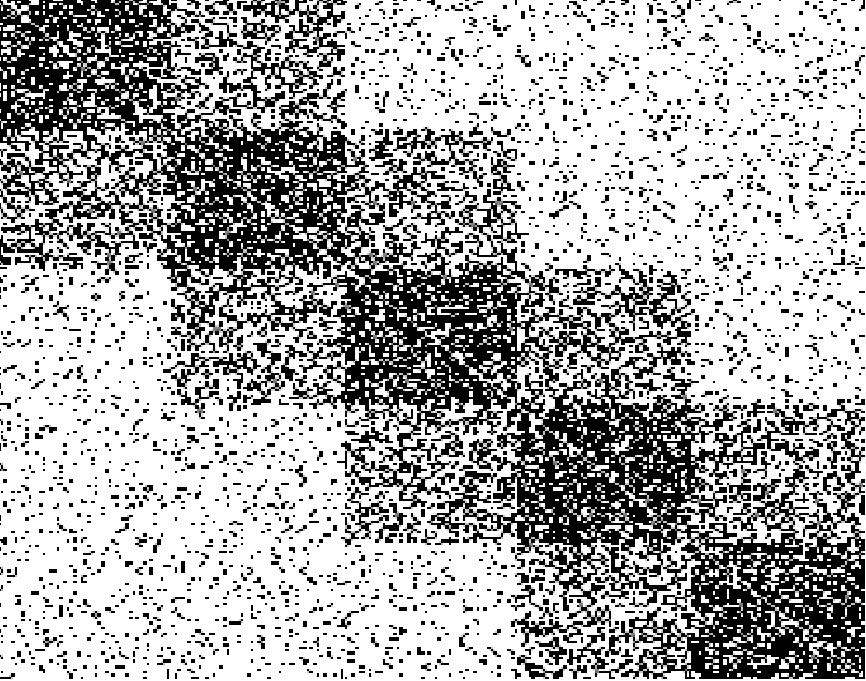}
&
\includegraphics[width=0.46\columnwidth]{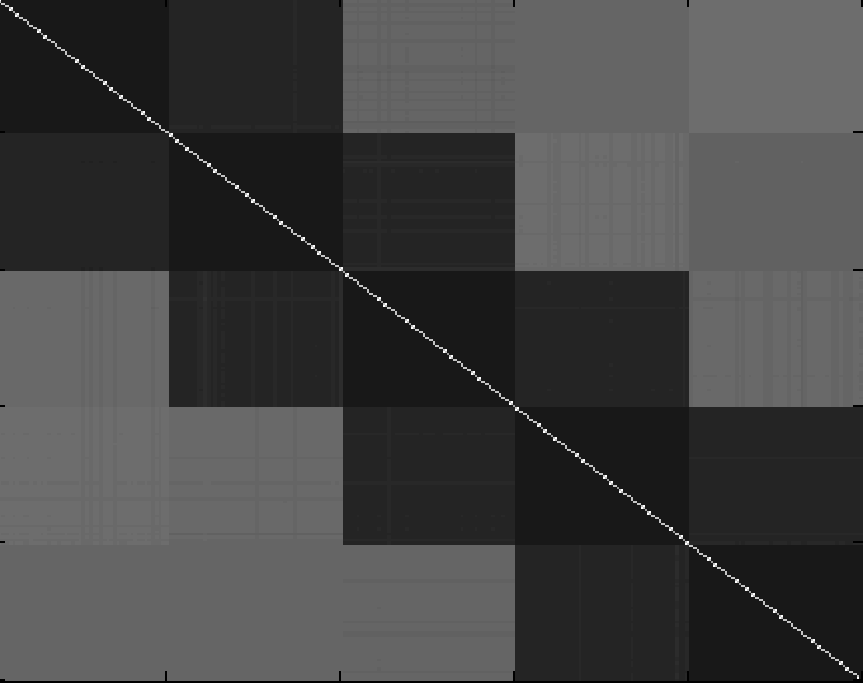}
\\
{\small (d.1)} & {\small (d.2)}
\\
\end{tabular}
\vspace{-12pt}
\caption{
Comparisons of ground-truth values and those inferred by
  CosTot. (a.1/2) The ground-truth/inferred distribution over topics for a
  particular community on the synthetic dataset.  (b.1/2) The ground-truth/inferred distribution over words
  for a particular topic.  (c.1/2) The ground-truth/inferred
  distribution over time stamps for a particular (topic, community)
  pair.  (d.1/2) The ground-truth adjacency matrix/inferred
  link probability matrix of the network. }
\label{fig:syn_data}
\end{figure}

{\bf Topic and community Generation.}\quad
We generate $K$ topics each associated with a multinomial
distribution over $V$ words obtained by discretizing
Gaussian distributions with means sampled uniformly on
$[0, V]$. One example of generated topic-word distribution is
shown in Figure~\ref{fig:syn_data}(b.1).
We then generate $C$ synthetic communities each associated
with a multinomial distribution over topics obtained by the
similar method to topic-word distribution but with the mean
sampled on $[0, K]$. Figure~\ref{fig:syn_data}(a.1) shows the
topic distribution
of one of the communities. For each (topic, community)
pair, we also use a discretized Gaussian distribution to
mimic the temporal variation
of the topic in the community, with the mean sampled
uniformly on $[0, T]$. Here $T$ is the number of time stamps.
One example of temporal variation is demonstrated in
Figure~\ref{fig:syn_data}(c.1), where we use smoothed curve
to fit the discrete values.

We also generate $U$ users, each of which is randomly
assigned a community label denoting the major community
that the user belongs to.

{\bf Text Generation.}\quad
Each user has $D_{i}$ posts. To mimic the property
of mixed membership in social media, the community that a
user belongs to when she creates the post is sampled from
a discretized Gaussian distribution over communities,
with the mean equal to her community label. After that,
topic, words and time stamp of the post are generated according
to the generative process described above (with background
topic omitted). Each post contains $W$ words.

{\bf Link Generation.}\quad
First we devise the following
link probability between two communities $i$ and $j$,
\begin{equation}\label{eq:link-probability}
\begin{split}
P_{ij} = \text{max}\{P_{0} - P_{slope} \cdot |i - j|, P_{min}\},
\end{split}
\end{equation}
where $i$ and $j\in\{1,\dots, C\}$ are the natural number indexes
of the communities; $P_{0}$ denotes the link probability within a
community (i.e. $P_{ij} = P_{0}\ \text{if}\ i = j$); $P_{slope}$
is the slope; and $P_{min}$ is the threshold minimum link
probability between two communities. Eq.(\ref{eq:link-probability})
implies that links are more likely generated between communities indexed by
consecutive numbers. Second, for each user pair, we find their
major communities of each and use the link probability between
those two communities to generate links between the user pair.
Figure~\ref{fig:syn_data}(d.1) shows the adjacency matrix of
users indicated by the generated links. Here rows (i.e. users)
are reordered according to their community labels. The resulting
block structure due to Eq.(\ref{eq:link-probability}) allows easy
comparison with the estimated link probabilities from our model.


{\bf Experimental Results.}\quad
We set $C = 5$, $K = 30$, $V = 100$, $T = 30$, $U = 250$, $D_i =
50$, $W = 20$, $P_{0} = 0.7$, $P_{slope} = 0.3$ and $P_{min} =
0.1$. All Gaussian variances are set to $1.0$.
We train CosTot
on this synthetic data, with the numbers of communities and
topics set to the true values.

Results are shown in the right column of
Figure~\ref{fig:syn_data}. We observe that the distributions
inferred by CosTot well match the ground truth.
Figure~\ref{fig:syn_data}(d.2) shows the matrix with each cell
representing the link formation probability between two users.
CosTot also well recovers the network structure.

\subsection{Experiments on Real-world Data}

We now present empirical results of our approach on Sina Weibo dataset (Section~\ref{sec:application}).   
We first describe the settings of experiments, then quantitatively evaluate our model in terms of three aspects: {\it time stamp prediction} for evaluating the capacity of capturing temporal dynamics, {\it link prediction} for measuring the capacity of modeling network, and {\it perplexity} for evaluating the capacity of modeling text. We only report the optimal results of different methods after tuning the parameters, and leave the study of parameter impacts for the end of the section.

\subsubsection{Experimental Settings}

\begin{figure}[!htp]
\centering
\includegraphics[width=0.95\columnwidth]{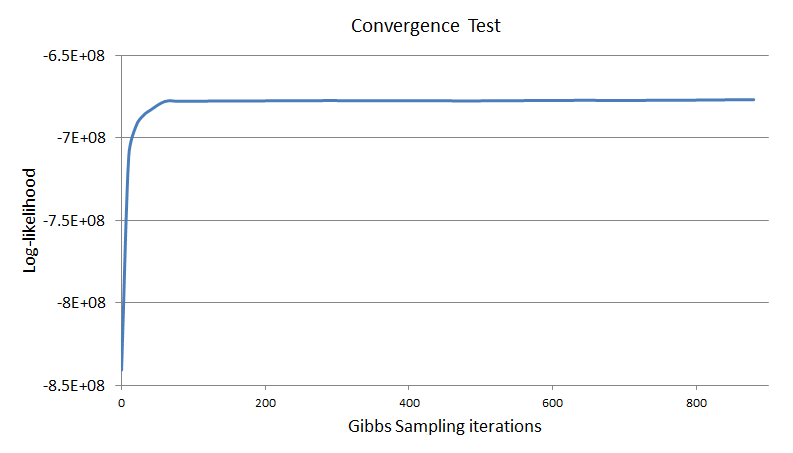}
\vspace{-10pt}
\caption{The complete log-likelihood of the model as a function
  of the number of Gibbs Sampling iterations.
The convergence is roughly the same under different model configurations.}
\label{fig:loglikelihood}
\end{figure}

We fix the hyperparameters to
$\rho=\alpha=\beta=\epsilon=\delta_{0}=0.01$ and $\delta_{1}=1$,
and $\lambda_{0}$ and $\lambda_{1}$ are determined as described
in Section~\ref{subsec:costot}. We calculate the {\it
complete} log-likelihood of the data as the proxy to monitor the
convergence of the Gibbs sampling algorithm,
\begin{equation*}
\begin{split}
 \mathcal{L} =&
   \text{log}\bigg(\prod_{i}\prod_{j}P(c_{ij}|\pi_{i})P(z_{ij}|\theta_{c_{ij}})
       P(t_{ij}|\varphi_{z_{ij}c_{ij}})\\
   &\ \ \ \ \ \ \ \ \ \prod_{l}P(w_{ijl}|\phi_{z_{ij}})^{f_{ijl}}P(w_{ijl}|\phi_{B})^{1-f_{ijl}}\bigg)\\
   &+\text{log}\bigg(\prod_{i}\prod_{\substack{i' \\ e_{ii'}\in\mathcal{E}_{i}}}P(s_{ii'}|\pi_{i})
     P(s'_{ii'}|\pi_{i'})P(e_{ii'}|\eta_{s_{ii'}s'_{ii'}})\bigg),
\end{split}
\end{equation*}
where the first part on the RHS is the log-likelihood of text and
time, and the second part is that of links.
Figure~\ref{fig:loglikelihood} shows that the log-likelihood as a
function of the number of iterations. It converges after a small
number of iterations. The convergence behavior is roughly the
same under different configurations of the number of communities
and the number of topics.  In particular, we set the number of
iterations to 500 in the following experiments. All our experiments are conducted on a Linux Server with eight 2.4GHz CPU cores and 32G memory.

\subsubsection{Time Stamp Prediction}

The task of time stamp prediction provides a way to quantitatively evaluate
the capacity of capturing temporal dynamics of
topics~\cite{wang2006topics}. We compare our proposed CosTot
with the following four competitors, where the first two
are the existing methods providing state-of-the-art performance on
this task; the third one corresponds to a subpart of CosTot; and the last one is an alternative approach of uncovering community-specific temporal variations.

{\bf Topics over Time (TOT).}\quad
Similar to CosTot, TOT~\cite{wang2006topics} jointly models the text and
time stamp of a document by treating both words and time
stamps as variables generated by latent topics. It employs a Beta
distribution to model the time distribution of each topic. TOT does
not exploits link data.

{\bf Enhanced User-Temporal Model with Burst-weighted Smoothing (EUTB).}\quad
EUTB~\cite{yin2013AUnified} incorporates time information by
assuming that a topic is generated either by a user or a time
stamp. Hence it models the topic distributions of users and
time stamps. Network data is exploited as a regularization based on the observation that neighbors in social network tend to have similar interests.
Note that an array of regularization methods are
proposed in \cite{yin2013AUnified}, while EUTB, with the link
regularization and burst-weighted smoothing, performs best
in the time stamp prediction task among a host of competitors.

{\bf CosTot without Link (CosTot-NoLink).}\quad
CosTot-NoLink is a subpart of CosTot with the network
component~(Section~\ref{subsec:costot}) removed. Hence
it provides a more fairly comparison with TOT.
We can also take a look at the impact of considering
link information by comparing to CosTot.

{\bf Simple Approach of Community-specific Temporal Variations (Simple-CTV).}\quad
One alternative method to analyze community-specific topic dynamics is to exploit network and text data step by step: we begin by dividing users into communities using well-established network community detection techniques, then uncover topic variation in certain community by running TOT on the posts generated by its members. By comparing with the simple approach, we gain insight into the benefits of combining the two aspects of community and topic in the CosTot way. In the experiment, we capture user multiple memberships by running MMSB on user network and assigning each user to two communities~\cite{xie2011overlapping} with highest probabilities.

We randomly select $20\%$ of the posts as the test set, while the
remaining $80\%$ posts and all links are used to train the
models. We set the hyperparameters in above baselines closely
resemble those in our model. Here we only report the best results of each with tuned parameters. The impact of model parameters is discussed in later section.
%

\begin{figure}[!htp]
\centering
\includegraphics[width=1.0\columnwidth]{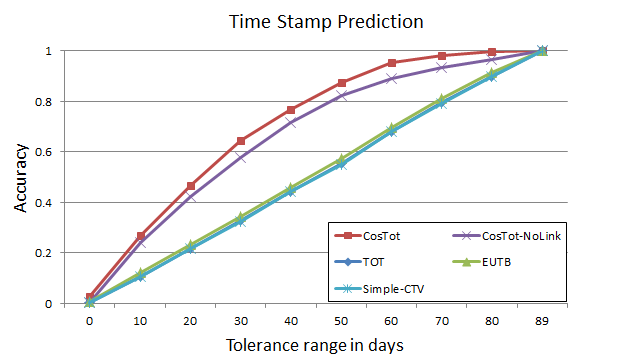}
\vspace{-13pt}
\caption{
  The prediction accuracy as a function of the tolerance range in days. For example, if we see that the difference
  between predicted time stamp and ground truth time stamp is within 10
  days as accurate, our model achieves accuracy of $26.7\%$ while
  TOT gives $10.7\%$. The best results are obtained by setting
  $K=100$ for all the four models, and $C=100$ for CosTot and
  CosTot-NoLink.}
\label{fig:accuracy}
\end{figure}

Figure~\ref{fig:accuracy} shows the prediction accuracy as a
function of tolerance range for these models. From the figure, we
see that our model performs better than all competitors.
Moreover, CosTot-NoLink outperforms TOT and EUTB, justifying the
advantage of distinguishing temporal variations of topics in
different communities, while the superiority of CosTot to
CosTot-NoLink shows the benefit brought by incorporating link
structures in social media.

We also observe that, Simple-CTV, despite taking into account community-specific topic dynamics, has poor performance as TOT. The reason is that it exploits network and content information separately, while ignores the correlations between them. Another drawback of Simple-CTV worth mentioning is that, since it runs TOT separately on different corpus, topics in a certain community are not shared by others. Therefore, it only provides us with disjoint views of different communities, and fails to consider the social media as a whole.

\subsubsection{Link Prediction}

Link prediction is a natural generalization task in networks, and a way to measure the quality of our model in modeling link structures. As discussed in Section~\ref{sec:app-pred}, we can predict links between users where the probability is above some threshold. However, since we are unaware of this threshold, we turn to {\it area under the receiver operating characteristic curve}~(AUC)~\cite{hanely1982roc} as the metric of the accuracy of prediction algorithm. Given the rank of all non-observed links, the AUC value can be interpreted as the probability that a randomly chosen true positive link is ranked above a randomly chosen true nonexistent link.

We compare CosTot with MMSB~\cite{airoldi2008mixed} and Link-PLSA-LDA~\cite{nallapati2008link}. MMSB exploits only the network data, while Link-PLSA-LDA incorporates both network and text information.

{\bf Mixed Membership Stochastic Blockmodel (MMSB).}\quad
Similar to CosTot, MMSB infers a probability distribution over communities for each user, and a link formation probability for each community pair. MMSB does not exploits text data.

{\bf Link-PLSA-LDA.}\quad
Link-PLSA-LDA defines a generative process for both text and citations between documents, where text generation is following the LDA approach, and citations are models as multinomial sampling of the target document from a topic-specific distribution over documents. We can also interpret Link-PLSA-LDA in another way by regarding documents as users in social media, words as user-generated text, and citations as directed links between users. In this perspective, links and text are generated by the same latent factor, which means one community is bound to one topic (as we see the latent factor as community when generating links, and as topic when generating text).

\begin{figure}[!htp]
\centering
\includegraphics[width=0.9\columnwidth]{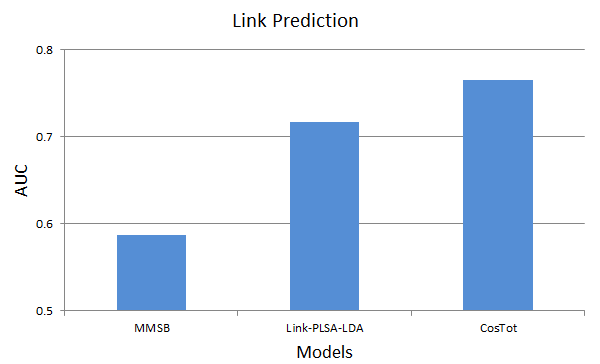}
\vspace{-8pt}
\caption{The AUC values of the link prediction task. The best results are obtained by setting $K = 100$ for Link-PLSA-LDA and CosTot, and $C = 100$ for MMSB and CosTot.
  }
\label{fig:link-auc}
\end{figure}

We randomly select $20\%$ of the positive links and $1\%$ of the negative links to evaluate the AUC; models are trained on the remaining links and all posts.
Figure~\ref{fig:link-auc} gives the AUC values for the three models. We see that by incorporating content information of users, Link-PLSA-LDA and CosTot outperform MMSB significantly.
CosTot outperforms Link-PLSA-LDA, since it is coherent with the fact that a community in the real world have varying levels of interests in multiple topics.

\subsubsection{Perplexity}

We evaluate the quality of our proposed CosTos in modeling text by computing the {\it perplexity}~\cite{blei2003latent} of a held-out test set. As a widely used metric in language modeling, perplexity monotonically decreases in the likelihood of the test data. A lower perplexity value indicates better generalization performance. For a test set of $M$ posts, the perplexity is:
\begin{equation*}
\begin{split}
    \text{\it perplexity}(\text{D}_{test}) =
      \exp\Big\{-\frac{\sum_{d=1}^{M}\log{p{({\bf w}_{d})}}}{\sum_{d=1}^{M}N_{d}}\Big\},
\end{split}
\end{equation*}
where $p{({\bf w}_{d})}$ is the probability of the words in the test post; for CosTot, it is computed as:
\begin{equation*}
\begin{split}
    p{({\bf w}_{d})} = &\sum_{c}P(c|\pi_{i})\sum_{k}P(k|\theta_{c})\\
&\prod_{l}(\chi P(w_{dl}|\phi_{k})
+(1-\chi)P(w_{dl}|\phi_{B})),
\end{split}
\end{equation*}
where $i$ is the author of the post.

\begin{figure}[!htp]
\centering
\includegraphics[width=0.9\columnwidth]{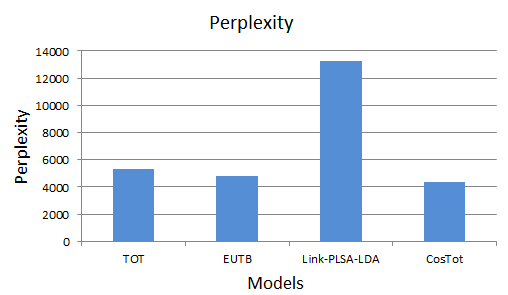}
\vspace{-8pt}
\caption{The perplexity values. The best results are obtained by setting $K = 100$ for all the four models, and $C = 100$ for CosTot.
  }
\label{fig:perplexity}
\end{figure}

We compare CosTot with three competitors: TOT, EUTB and Link-PLSA-LDA, and the results are shown in Figure~\ref{fig:perplexity}. We see that CosTot has the lowest perplexity (i.e. best text prediction performance) among all the competitors. In contrast, Link-PLSA-LDA shows a poor performance, since its topics are tangled with communities in the same latent factor, and thus their fitness in modeling text is weakened by links.

\subsubsection{Parameter Study}

The two parameters, i.e., the number of communities $C$ and the number of topics $K$, are critical to the performance of CosTot. We therefore study the impacts of these parameters in different tasks. Here we show the experimental results for link prediction and perplexity, which demonstrate the different roles of $C$ and $K$ in determining the model performance.

\begin{figure}[t!]
\centering
\includegraphics[width=0.9\columnwidth]{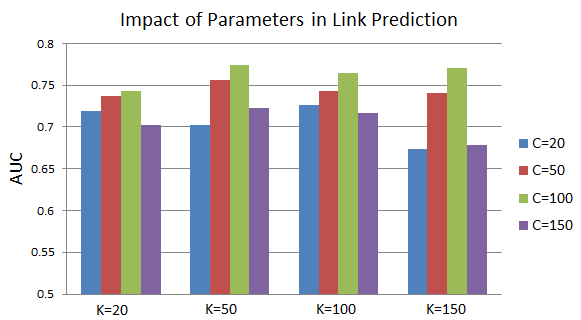}
\vspace{-8pt}
\caption{The impact of model parameters $C$ and $K$ in the task of link prediction.
  }
\label{fig:C-K-link-AUC}
\end{figure}

Figure~\ref{fig:C-K-link-AUC} shows the AUC values of link prediction under different settings, $C \in \{20,50,100,150\}$ and $K \in \{20,50,100,150\}$. We see that given a fixed $K$, the AUC value at first increases as $C$ increases, and there is an intermediate value of $C$ (i.e. 100) at which CosTot has the best performance. After that the AUC value decreases as $C$ continues to increase. On the other hand, given any fixed $C$, the result fluctuates slightly without a clear pattern as $K$ varies, indicating that the number of topics is less important for link prediction than the number of communities. The underlying reason is that, in CosTot links are generated by mixture of communities, hence the number of communities directly impacts the capacity of modeling network. In contrast, although there exists correlations between text and network, the influence of topics on network modeling is indirect.

\begin{figure}[t!]
\centering
\includegraphics[width=0.9\columnwidth]{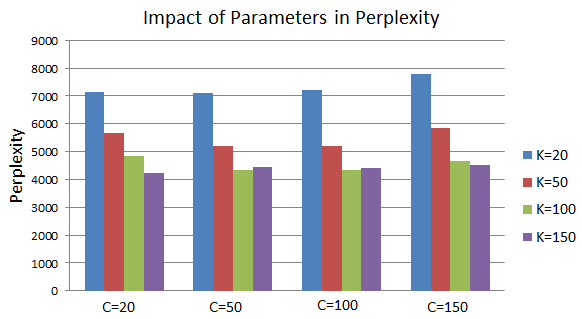}
\vspace{-8pt}
\caption{The impact of model parameters $C$ and $K$ in perplexity.
  }
\label{fig:C-K-perplexity}
\end{figure}

Figure~\ref{fig:C-K-perplexity} shows the impacts of $C$ and $K$ with regard to the quality of CosTot in text modeling. We see that perplexity decreases with the increasing number of topics, while remain stable as the number of communities varies. The result is reasonable since topics account for generating text. It is also worth mentioning that the performance does not change significantly when $K$ is larger than 100.

\section{Conclusion}

In this paper, we have addressed the problem of temporal topic dynamics within different
communities in social media. We presented CosTot (Community Specific
Topics-over-Time), a probabilistic longitudinal model jointly
over network, text and time, to simultaneously uncover the hidden
topics and communities, and capture the community-specific temporal
variation of topics.
We provided efficient inference implementation and abundant applications to demonstrate the feasibility and usefulness of this model. In the empirical study, our model achieved best performance in the tasks of time stamp prediction, link prediction
and text perplexity among several competitors. We also provided several novel
visualization examples of topic temporal patterns at different
granularities, which clearly show how topics attract attentions from
different communities.


For future work, we are interested in extending the model to further
capture network dynamics. We also would like to
incorporate information diffusion among different communities that leads
to the observed temporal patterns.

\bibliographystyle{abbrv}
\small
\bibliography{hdpcommtempo}
\end{document}